\documentstyle[twoside,psfig]{article}

\catcode`\@=11
\long\def\@makefntext#1{
\protect\noindent \hbox to 3.2pt {\hskip-.9pt
$^{{\eightrm\@thefnmark}}$\hfil}#1\hfill}       

\def\@makefnmark{\hbox to 0pt{$^{\@thefnmark}$\hss}}    

\def\ps@myheadings{\let\@mkboth\@gobbletwo
\def\@oddhead{\hbox{}
\rightmark\hfil\eightrm\thepage}
\def\@oddfoot{}\def\@evenhead{\eightrm\thepage\hfil
\leftmark\hbox{}}\def\@evenfoot{}
\def\sectionmark##1{}\def\subsectionmark##1{}}

\oddsidemargin=\evensidemargin
\newcounter{sectionc}\newcounter{subsectionc}\newcounter{subsubsectionc}
\renewcommand{\section}[1] {\vspace{12pt}\addtocounter{sectionc}{1}
\setcounter{subsectionc}{0}\setcounter{subsubsectionc}{0}\noindent
    {\tenbf\thesectionc. #1}\par\vspace{5pt}}
\renewcommand{\subsection}[1] {\vspace{12pt}\addtocounter{subsectionc}{1}
    \setcounter{subsubsectionc}{0}\noindent
    {\bf\thesectionc.\thesubsectionc. {\kern1pt \bfit #1}}\par\vspace{5pt}}
\renewcommand{\subsubsection}[1] {\vspace{12pt}\addtocounter{subsubsectionc}{1}
    \noindent{\tenrm\thesectionc.\thesubsectionc.\thesubsubsectionc.
    {\kern1pt \tenit #1}}\par\vspace{5pt}}
\newcommand{\nonumsection}[1] {\vspace{12pt}\noindent{\tenbf #1}
    \par\vspace{5pt}}

\newcounter{appendixc}
\newcounter{subappendixc}[appendixc]
\newcounter{subsubappendixc}[subappendixc]
\renewcommand{\thesubappendixc}{\Alph{appendixc}.\arabic{subappendixc}}
\renewcommand{\thesubsubappendixc}
    {\Alph{appendixc}.\arabic{subappendixc}.\arabic{subsubappendixc}}

\renewcommand{\appendix}[1] {\vspace{12pt}
        \refstepcounter{appendixc}
        \setcounter{figure}{0}
        \setcounter{table}{0}
        \setcounter{lemma}{0}
        \setcounter{theorem}{0}
        \setcounter{corollary}{0}
        \setcounter{definition}{0}
        \setcounter{equation}{0}
        \renewcommand{\thefigure}{\Alph{appendixc}.\arabic{figure}}
        \renewcommand{\thetable}{\Alph{appendixc}.\arabic{table}}
        \renewcommand{\theappendixc}{\Alph{appendixc}}
        \renewcommand{\thelemma}{\Alph{appendixc}.\arabic{lemma}}
        \renewcommand{\thetheorem}{\Alph{appendixc}.\arabic{theorem}}
        \renewcommand{\thedefinition}{\Alph{appendixc}.\arabic{definition}}
        \renewcommand{\thecorollary}{\Alph{appendixc}.\arabic{corollary}}
        \noindent{\tenbf Appendix \theappendixc #1}\par\vspace{5pt}}
\newcommand{\subappendix}[1] {\vspace{12pt}
        \refstepcounter{subappendixc}
        \noindent{\bf Appendix \thesubappendixc. {\kern1pt \bfit #1}}
    \par\vspace{5pt}}
\newcommand{\subsubappendix}[1] {\vspace{12pt}
        \refstepcounter{subsubappendixc}
        \noindent{\rm Appendix \thesubsubappendixc. {\kern1pt \tenit #1}}
    \par\vspace{5pt}}

\newcommand{\textlineskip}{\baselineskip=13pt}
\newcommand{\smalllineskip}{\baselineskip=10pt}

\def\eightcirc{
\begin{picture}(0,0)
\put(4.4,1.8){\circle{6.5}}
\end{picture}}
\def\eightcopyright{\eightcirc\kern2.7pt\hbox{\eightrm c}}

\newcommand{\copyrightheading}[1]
    {\vspace*{-2.5cm}\smalllineskip{\flushleft
    {\footnotesize International Journal of Modern Physics C #1}\\
    {\footnotesize $\eightcopyright$\, World Scientific Publishing
     Company}\\
     }}

\def\abstracts#1#2#3{{
    \centering{\begin{minipage}{4.5in}\footnotesize\baselineskip=10pt
    \parindent=0pt #1\par
    \parindent=15pt #2\par
    \parindent=15pt #3
    \end{minipage}}\par}}

\def\keywords#1{{
    \centering{\begin{minipage}{4.5in}\footnotesize\baselineskip=10pt
    {\footnotesize\it Keywords}\/: #1
    \end{minipage}}\par}}

\newcommand{\bibit}{\nineit}
\newcommand{\bibbf}{\ninebf}
\renewenvironment{thebibliography}[1]
        {\frenchspacing
     \ninerm\baselineskip=11pt
         \begin{list}{\arabic{enumi}.}
        {\usecounter{enumi}\setlength{\parsep}{0pt}
     \setlength{\leftmargin 12.7pt}{\rightmargin 0pt} 
         \setlength{\itemsep}{0pt} \settowidth
    {\labelwidth}{#1.}\sloppy}}{\end{list}}

\newcounter{itemlistc}
\newcounter{romanlistc}
\newcounter{alphlistc}
\newcounter{arabiclistc}

\newcommand{\fcaption}[1]{
        \refstepcounter{figure}
        \setbox\@tempboxa = \hbox{\footnotesize Fig.~\thefigure. #1}
        \ifdim \wd\@tempboxa > 5in
           {\begin{center}
        \parbox{5in}{\footnotesize\smalllineskip Fig.~\thefigure. #1}
            \end{center}}
        \else
             {\begin{center}
             {\footnotesize Fig.~\thefigure. #1}
              \end{center}}
        \fi}

\newcommand{\tcaption}[1]{
        \refstepcounter{table}
        \setbox\@tempboxa = \hbox{\footnotesize Table~\thetable. #1}
        \ifdim \wd\@tempboxa > 5in
           {\begin{center}
        \parbox{5in}{\footnotesize\smalllineskip Table~\thetable. #1}
            \end{center}}
        \else
             {\begin{center}
             {\footnotesize Table~\thetable. #1}
              \end{center}}
        \fi}

\def\@citex[#1]#2{\if@filesw\immediate\write\@auxout
    {\string\citation{#2}}\fi
\def\@citea{}\@cite{\@for\@citeb:=#2\do
    {\@citea\def\@citea{,}\@ifundefined
    {b@\@citeb}{{\bf ?}\@warning
    {Citation `\@citeb' on page \thepage \space undefined}}
    {\csname b@\@citeb\endcsname}}}{#1}}

\newif\if@cghi
\def\cite{\@cghitrue\@ifnextchar [{\@tempswatrue
    \@citex}{\@tempswafalse\@citex[]}}
\def\citelow{\@cghifalse\@ifnextchar [{\@tempswatrue
    \@citex}{\@tempswafalse\@citex[]}}
\def\@cite#1#2{{$\null^{#1}$\if@tempswa\typeout
    {IJCGA warning: optional citation argument
    ignored: `#2'} \fi}}

\def\pmb#1{\setbox0=\hbox{#1}
    \kern-.025em\copy0\kern-\wd0
    \kern.05em\copy0\kern-\wd0
    \kern-.025em\raise.0433em\box0}

\def\fnt#1#2{\footnotetext{\kern-.3em
    {$^{\mbox{\scriptsize #1}}$}{#2}}}

\def\ps@myheadings{%
    \let\@oddfoot\@empty\let\@evenfoot\@empty
    \def\@evenhead{\slshape\leftmark\hfil}
    \def\@oddhead{\hfil{\slshape\rightmark}}
    \let\@mkboth\@gobbletwo
    \let\sectionmark\@gobble
    \let\subsectionmark\@gobble
    }
\font\tenrm=cmr10
\font\tenit=cmti10
\font\tenbf=cmbx10
\font\bfit=cmbxti10 at 10pt
\font\ninerm=cmr9
\font\nineit=cmti9
\font\ninebf=cmbx9
\font\eightrm=cmr8

\def\qed{\hbox{${\vcenter{\vbox{            
   \hrule height 0.4pt\hbox{\vrule width 0.4pt height 6pt
   \kern5pt\vrule width 0.4pt}\hrule height 0.4pt}}}$}}

\def\bsc{{\sc a\kern-6.4pt\sc a\kern-6.4pt\sc a}}   
\def\bflatex{\bf L\kern-.30em\raise.3ex\hbox{\bsc}\kern-.14em
T\kern-.1667em\lower.7ex\hbox{E}\kern-.125em X}

\begin{document}
\setlength{\textheight}{7.7truein}  

\thispagestyle{empty}

\markboth{\protect{\footnotesize\it T. Gao, F. L. Yan and Z. X. Wang}}{\protect{\footnotesize\it  Quantum Secure
Conditional Direct Communication via EPR  Pairs}}

\normalsize\textlineskip

\setcounter{page}{1}

\copyrightheading{}         

\vspace*{0.88truein}

\centerline{\bf  QUANTUM SECURE CONDITIONAL DIRECT  } \vspace*{0.035truein} \centerline{\bf COMMUNICATION VIA
EPR PAIRS}
 \vspace*{0.37truein}
\centerline{\footnotesize TING GAO} \baselineskip=12pt \centerline{\footnotesize\it College of Mathematics and
Information Science, Hebei Normal University} \baselineskip=10pt \centerline{\footnotesize\it Shijiazhuang,
050016, P. R. China}\centerline{\footnotesize\it CCAST (World Laboratory), P. O. Box 8730, Beijing, 100080, P.
R. China}\baselineskip=10pt \centerline{\footnotesize\it Department of Mathematics, Capital Normal University,
Beijing, 100037, P. R. China}
 \centerline{\footnotesize\it E-mail: gaoting@heinfo.net}

\vspace*{15pt}          
\centerline{\footnotesize FENGLI YAN} \baselineskip=12pt \centerline{\footnotesize\it  College of Physics, Hebei
Normal University, Shijiazhuang, 050016, China} \baselineskip=10pt \centerline{\footnotesize\it CCAST (World
Laboratory), P. O. Box 8730, Beijing, 100080, P. R. China} \centerline{\footnotesize\it E-mail:
flyan@mail.hebtu.edu.cn}

\vspace*{15pt}          
\centerline{\footnotesize ZHIXI WANG} \baselineskip=12pt \centerline{\footnotesize\it  Department of
Mathematics, Capital Normal University, Beijing, 100037, P. R. China} \centerline{\footnotesize\it E-mail:
wangzhx@mail.cnu.edu.cn} 

\vspace*{0.25truein} \abstracts{Two schemes for  quantum secure conditional direct communication are proposed,
where a set of
   EPR pairs  of maximally entangled particles in Bell states,  initially made by the
supervisor Charlie, but shared by the sender Alice and
    the receiver Bob,  functions as  quantum information channels for faithful transmission. After
insuring the security of the quantum channel and obtaining the permission of Charlie (i.e. Charlie is
trustworthy and cooperative, which means the 'conditional' in the two schemes), Alice and Bob begin their
private communication under the control of Charlie. In the first scheme, Alice transmits secret message to Bob
in a deterministic manner with the help of Charlie by means of Alice's local unitary transformations, both Alice
and Bob's local measurements, and both of Alice and Charlie's public classical communication.
  In the second scheme, the secure communication
between Alice and Bob can be achieved  via public classical communication of Charlie and Alice, and the local
measurements of  both Alice and Bob. The common feature of these protocols is that the communications between
two communication parties Alice and Bob depend on the agreement of the third side Charlie. Moreover,
transmitting one bit secret message,  the sender Alice only needs to apply a local operation on her one qubit
and send one bit classical information.
 We also show that the two schemes are completely secure if
quantum channels are perfect.}{}{}

\vspace*{5pt} \keywords{ Quantum secure conditional direct communication; EPR pairs; local measurement; local
unitary transformation}


\vspace*{1pt}\textlineskip  
\section{Introduction}     
\vspace*{-0.5pt} \noindent In the information society, more and more people wish to transmit secret messages.
The way to achieve communication in secret is to use a cryptographic protocol, that is, before transmitting
their secret messages the two distant communication parties must distribute secret key first. Although in many
ways, the secret key distribution problem is just as difficult as the original problem of communication in
private through a classical channel, fortunately, quantum mechanics can be used to do key distribution in such a
way that Alice and Bob's security can not be compromised. This procedure is known as quantum key distribution or
quantum cryptography. In 1984, Bennett and Brassard presented the first quantum cryptography,  BB84
protocol,$^1$  using quantum mechanics to distribute keys between Alice and Bob, without any possibility of a
compromise.
 Since then Ekert proposed a quantum key distribution scheme depending on the correlation of Einstein-Podolsky-Rosen
(EPR) pair, the maximally entangled state of two particles in 1991,$^2$  Bennett   put forward a quantum
cryptography scheme known as B92 protocol,$^3$  and  numerous  quantum cryptographic protocols on both
theoretical and experimental aspects$^{[4-19]}$  have been proposed.

Shimizu and Imoto$^{20,21}$ and Beige et al.$^{22}$ presented  novel quantum secure direct communication
 schemes, in which the two parties   communicate important messages directly without first establishing a
shared secret key to encrypt them and the message is deterministically sent through the quantum channel, but can
be read only after a final transmission of an additional  classical information for each qubit. Thereafter, many
quantum secure direct communication  schemes, such as Bostr\"{o}m and Felbinger's$^{23}$ "ping pong protocol"
which is insecure if it is operated in a noisy quantum channel, as indicated by W$\acute{\rm o}$jcik,$^{24}$ and
Deng et al.'s  two schemes, one  using EPR pair block$^{25}$ and  the other   with a quantum
one-time-pad,$^{26}$ were proposed. Unfortunately,  in all these secure direct communication protocols it is
necessary to send the qubits carrying secret messages in the public channel. Therefore a potential eavesdropper,
Eve, can intercept the qubits in transmission and make the communication interrupted. In order to prevent the
qubits transmitted in the public channel, Yan et al. put forward three protocols for quantum secure direct
communication, using EPR pairs and teleportation,$^{27}$  by EPR pairs and entanglement swapping,$^{28}$  and
based on GHZ states and entanglement swapping,$^{29}$ respectively. We suggested two schemes for controlled and
secure direct communication using three-particle entangled state and teleportation.$^{30,31}$  Cai gave an
one-time-pad key communication protocol with entanglement.$^{32}$  Since there is not a transmission of the
qubits with the secret messages between two communication parties in the public channel, they are secure for
secret direct communication if perfect quantum channel is used. However, in the protocols of Refs. [27, 30, 31],
for transmitting one bit secret message, Alice has to make Bell measurement on two qubit and send two bits
classical information to Bob because of   four random outcomes produced by Bell measurement, so it would waste
the classical information resource. In this paper, we would like to simplify these protocols  and give two even
simpler but more economical and more efficient schemes with EPR pairs.  In these schemes, for her to transmit
one bit secret message, Alice only needs to perform a local operation on one qubit and send one bit classical
information to Bob. Therefore, our proposed schemes are easy to realize in experiment and greatly economize
classical information resource.

\section{Quantum Secure Conditional Direct Communication via EPR Pairs}
\noindent Consider the following scenario. Suppose that the administrative personnel of the server,  Charlie,
wishes to control the correspondence between users. This means that if and only if Charlie gives  his permission
(that is, Charlie is trustworthy and cooperative), Alice and Bob can correspond with each other. However, the
users wish to communicate in secret and not altered during transmission. What can they do? Our following two
schemes will suit this task. In our schemes, there are three parties: the controller Charlie and the users Alice
and Bob.  The present two  communication protocols, called quantum secure conditional direct communication
protocols, are designed to work only provided Charlie wants them to do so. Here the word 'conditional' means
that Charlie is trustworthy and cooperative.  How do these new schemes work?

First, the controller Charlie  prepares EPR pairs in Bell states:
\begin{eqnarray}\label{EPR}
\Phi^+ & \equiv & \frac{1}{\sqrt{2}}(|00\rangle+|11\rangle) \nonumber\\
       & = & \frac{1}{\sqrt{2}}(|++\rangle+|--\rangle)\nonumber\\
       &=& \frac{1}{\sqrt{2}}(|+y\rangle|-y\rangle+|-y\rangle|+y\rangle),\label{EPR1}\\
\Phi^- & \equiv & \frac{1}{\sqrt{2}}(|00\rangle-|11\rangle)\nonumber\\
       & = & \frac{1}{\sqrt{2}}(|+-\rangle+|-+\rangle)\nonumber\\
       &=& \frac{1}{\sqrt{2}}(|+y\rangle|+y\rangle+|-y\rangle|-y\rangle),\label{EPR2}\\
\Psi^+ & \equiv & \frac{1}{\sqrt{2}}(|01\rangle+|10\rangle)\nonumber\\
       & = & \frac{1}{\sqrt{2}}(|++\rangle-|--\rangle)\nonumber\\
       &=& \frac{i}{\sqrt{2}}(|-y\rangle|-y\rangle-|+y\rangle|+y\rangle),\label{EPR3}\\
\Psi^- & \equiv & \frac{1}{\sqrt{2}}(|01\rangle-|10\rangle)\nonumber\\
       & = & \frac{1}{\sqrt{2}}(|-+\rangle-|+-\rangle)\nonumber\\
       &=& \frac{i}{\sqrt{2}}(|+y\rangle|-y\rangle-|-y\rangle|+y\rangle),\label{EPR4}
\end{eqnarray}
and then sends one qubit of each pair to Alice and Bob. Here $|+\rangle=\frac{|0\rangle+|1\rangle}{\sqrt{2}}$,
$|-\rangle=\frac{|0\rangle-|1\rangle}{\sqrt{2}}$, $|+y\rangle=\frac{|0\rangle+i|1\rangle}{\sqrt{2}}$, and
$|-y\rangle=\frac{|0\rangle-i|1\rangle}{\sqrt{2}}$. These EPR pairs, shared by the sender Alice and the receiver
Bob, function as quantum information channel.  In order to keep perfect quantum channel, Alice, Bob and Charlie
then choose randomly a subset of EPR pairs,
 and do   some  appropriate tests of fidelity. They can use the schemes testing the security of EPR pairs
(quantum channel) in Refs.[2, 4, 13, 25, 27]. Passing the test insures that they continue to hold sufficiently
pure, maximally entangled quantum states (i.e. the quantum channel
 is perfect). However, if tampering has occurred, they discard these EPR pairs and
construct new EPR pairs again.

After insuring the security of quantum channel and getting the approval of Charlie (this means that Charlie is
trustworthy and cooperative), Alice and Bob start their  quantum secure conditional direct communication under
the control of Charlie.

\subsection{Based on Local Unitary Transformations}
\noindent A.1 Alice, Bob and Charlie arrange all EPR pairs in random order.  The three of them agree on that the
four Bell states are encoded information of two classical bits
\begin{eqnarray}\label{EPRcode}
 \Phi^+\rightarrow 00, \Phi^-\rightarrow 01, \Psi^+\rightarrow 10, \Psi^-\rightarrow 11.
\end{eqnarray}

A.2 Alice and  Bob agree on  that Alice encodes information by  local operations
\begin{eqnarray}\label{operation}
\sigma_{0}=I=|0\rangle\langle 0|+|1\rangle\langle1|, \sigma_{1}=\sigma_z=|0\rangle\langle0|-|1\rangle\langle1|
\end{eqnarray}
 on EPR pairs, and  assign secretely one bit  to Alice's  operations
 as following encoding
\begin{equation}\label{m}
 \sigma_0\rightarrow 0,~~ \sigma_1\rightarrow 1.
\end{equation}

A.3 Alice  encodes her secret messages (secret classical bits) on  the  EPR pairs sequence. Explicitly, Alice
applies a predetermined unitary operation on each of her particles   according to her  secret message sequence.

 Suppose that Alice and Bob initially share EPR pair $\Phi^+$, and
 Alice wants to transmit 1 to Bob,  then Alice  performs a local
operation $\sigma_1$ on her particle, thus  the
   state  $\Phi^+$ is turned into  $\Phi^-$.

A.4  Alice performs a local measurement on her qubit, in the  basis $\{|+\rangle, |-\rangle\}$, and classically
(and publicly) broadcasts the measurement outcome (on which basis state of $\{|+\rangle, |-\rangle\}$ he
obtained by projection) to the receiver Bob.

A.5 Bob makes a local measurement on his particle also in the basis $\{|+\rangle, |-\rangle\}$. After that he
 informs Charlie that he has made a  measurement on his qubit over a classical
channel, but does not tell the result of his measurement.

From Eq.(\ref{EPR2}), Bob's possible post-measurement states of his particle are  $|-\rangle$ and $|+\rangle$
depending on Alice's possible  measurement outcomes $|+\rangle$ and $|-\rangle$, respectively.

A.6 Charlie tells  Bob which  EPR pair Alice and Bob initially share. To do this, he needs to inform Bob two
classical bits through classical channel.

A.7 Bob can read out Alice's secret message by Charlie's classical information and  the two measurement results
of both Alice and himself.

From Charlie's two bits classical information 00, Bob knows  that Alice and he initially share EPR pair
$\Phi^+$. By Eq.(\ref{EPR1}) and  both Alice and
 Bob's outcomes of measurement, Bob can deduce Alice's secret message 1.

In this scheme,  EPR pairs functioning as quantum channel must be at least in two Bell states. If only EPR pairs
in one Bell state function as quantum channel, then Charlie have no control over the two communication parties
Alice and Bob, since Bob can infer Alice's secret message only by Alice and Bob's measurement outcomes.

\subsection{Based on Local Measurements}
\noindent B.1  Alice, Bob and Charlie randomly sequence all EPR pairs  and agree beforehand as the encoding
Eq.(\ref{EPRcode}).

B.2 Alice and Bob also agree on that if Alice's measurement outcome is the same with the
 secret message to be transmitted, then Alice sends classical information 0 to Bob, otherwise she sends
 classical information 1 to Bob.

B.3 Alice makes a local measurement on each of  her particles, in the computational basis $\{|0\rangle,
|1\rangle\}$, and publicly sends Bob the corresponding classical information according to the secret message to
be transmitted to Bob over classical channel. Explicitly, if Alice's measurement outcome is $|0\rangle$ and the
secret message is 0, or outcome $|1\rangle$ and secret message 1, then Alice sends classical information 0 to
Bob; otherwise, that is, Alice's result of measurement is $|0\rangle$ but her secret message is 1, or result
$|1\rangle$ but secret message 0, she transfers
 classical information 1 to Bob.

B.4 Bob performs a measurement on his qubit  along $z$-axis and makes known to Charlie, but does not announce
his measurement outcome.

B.5 Charlie informs  Bob which  EPR pair Alice and Bob initially share.

B.6 By his  outcomes of measurement and
 Charlie's classical information, Bob can deduce Alice's measurement  results.
 According to  the classical information he received from Alice, Bob can infer the secret message
  that Alice wants
 to transmit to him.

For instance, if Charlie's classical information is 0001101100011110, then the EPR pairs are $\Phi^+$, $\Phi^-$,
$\Psi^+$, $\Psi^-$, $\Phi^+$, $\Phi^-$,  $\Psi^-$ and $\Psi^+$, respectively. If Bob's measurement outcomes on
the eight EPR pairs  are the states $|0\rangle, |0\rangle,
 |1\rangle,|0\rangle,|1\rangle,|0\rangle,|0\rangle,|1\rangle$, then he can infer that Alice's outcomes of
 measurement are $|0\rangle, |0\rangle,
 |0\rangle,|1\rangle,|1\rangle,|0\rangle,|1\rangle,|0\rangle$. Based on Alice's  classical information
 10001111, Bob can deduce  the secret message 10010101 that Alice wishes to send to him.

In this scheme,  EPR pairs functioning as quantum channel must be at least in more than two Bell states. The
reason is as follows: without loss of generality, we suppose that only EPR pairs in the Bell states $\Phi^-$ and
$\Psi^+$ function as quantum channel. From Eqs.(\ref{EPR2})-(\ref{EPR3}),  Alice and Bob can get out of control
by means of measuring their respective particles in the same basis $\{|+y\rangle, |-y\rangle\}$.

 The common feature of the two schemes is that communication between two sides depends on the agreement of the
third side Charlie (i.e. Charlie is trustworthy and cooperative). If the controller Charlie does not transmit
classical information to Bob, that is, he does not tell Bob the real order of all EPR pairs (i.e. he is not
faithful), then Bob can neither infer Alice's local unitary operators in the first scheme nor her measurement
outcomes in the second scheme, therefore Bob can not obtain her secret messages. That is why we suppose that
Charlie is trustworthy and cooperative and call the two proposed protocols quantum secure conditional direct
communication protocols. The two schemes have the advantage of the schemes in Refs.[27, 30, 31]. For
transmitting one bit secret message, in Refs.[27, 30, 31], Alice has to make Bell measurement on two qubits and
send two bits classical information to Bob because of four random outcomes produced by Bell measurement, so it
would waste the classical information resource. However, in the above two schemes, Alice only needs to perform
two local operations (a local unitary operation and a local measurement) for the first scheme, or one local
operation (a local measurement) for the second scheme on one qubit and send one bit classical information to
Bob. Therefore, our proposed schemes are easy to realize in experiment and greatly save classical information
resource.

 At the same time these two quantum communication schemes
  can achieve faithfully transmission of delicate information across a
 noisy environment assuming that classical information is robust and easy to protect against noise (as it is).
 Also the entanglement of EPR pair is independent of the spatial location of Bob relative to Alice so that
 Alice and Charlie can transfer  information without even knowing Bob's location - they needs only to broadcast
 their
 encoding information via classical channel, for example Alice and Charlie can write their encoding information
 on a newspaper.
 Because the local
 measurement outcomes are random, so in these two protocols, the secret messages are transmitted without revealing any
 information to a potential eavesdropper if the quantum channel is perfect EPR pairs.

  \section{Security in the Proposed Protocols}

\noindent
 It is evident that  the security of these protocols
  are only determined by the quantum channel. If the quantum
 channel is perfect EPR pairs, the protocols must be  safety ones. By using the schemes testing the security of
 EPR pairs, we can obtain a perfect quantum channel. For example, Alice, Bob and Charlie choose randomly a
 subset of EPR pairs, then Alice and Bob perform local measurements on their respective qubits along $z$-axis or
 $x$-axis at random. After measurements, Charlie broadcasts the order of EPR pairs in the chosen subset.
 Depending on Charlie's classical information and comparing Alice and Bob's measurement outcomes, they can
 test the security of  EPR pairs and obtain a perfect quantum channel as in Ref.[27].
  Local measurement outcomes ensure the security of
  the two proposed schemes. Based on outcomes, Alice, Bob and
  Charlie verify whether there is an attack by Eve and obtain a perfect quantum channel.
 Therefore these two schemes for quantum direct communication using
 EPR pairs and local operations are absolutely reliable, deterministic and secure.

How is it guaranteed that the two  quantum secure conditional direct communications are secure against attacks
by an eavesdropper? An eavesdropper, Eve, may have several kinds of strategies to attack. We shall consider that
Eve enforces the intercept-resend strategy and tries to extract some information. We show that authorized
persons can detect these attacks as they examine the quantum channel.

As mentioned in Ref. [27],  Eve can  obtain information if the quantum channel is not the perfect EPR pairs.

For example if Eve couples EPR pair with her probe in preparing or distributing EPR pair, and makes the quantum
channel in GHZ states
\begin{eqnarray}\label{GHZ}
    &|P^{\pm}\rangle\equiv\frac{1}{\sqrt{2}}(|000\rangle\pm|111\rangle),
    |Q^{\pm}\rangle\equiv\frac{1}{\sqrt{2}}(|001\rangle\pm|110\rangle),\nonumber\\
    &|R^{\pm}\rangle\equiv\frac{1}{\sqrt{2}}(|010\rangle\pm|101\rangle),
    |S^{\pm}\rangle\equiv\frac{1}{\sqrt{2}}(|011\rangle\pm|100\rangle),
\end{eqnarray}
then she is in the same position with the legitimate party Alice or Bob. But this case can be ruled out as long
as Alice and Bob check the EPR pairs by means of their  local measurement in the basis $\{|0\rangle,
|1\rangle\}$ or basis $\{|+\rangle, |-\rangle\}$ randomly and comparing their measurement results as stated in
Ref.[27]. If Eve uses so called entanglement pair method  to obtain information, she will also be found by Alice
and Bob's test, which was shown in Ref.[27]. So in any case, as long as an eavesdropper exists, we can find her
and insure the security of quantum channel to realize secure quantum communication.

In summary we provide two  schemes for quantum secure conditional direct communication via EPR pairs,  one based
on local unitary operations  and local measurements, the other by  local measurements. In the two protocols, a
set of EPR pairs of maximally entangled particles in Bell states, initially made by the supervisor Charlie, but
shared by the sender Alice and
    the receiver Bob,  functions as  quantum information channels for faithful transmission. After
insuring the security of the quantum channels, if Charlie would like to help Alice  and Bob to communicate, that
is, Charlie is trustworthy and cooperative, which is the condition of the two schemes, Alice and Bob can
communicate important messages directly under the control of the third side Charlie. In the first scheme,  Alice
applies local unitary operators and local measurement on her particles, and Bob performs local measurements on
his portion. Bob can deduce Alice's secret message  by Charlie's public communication,
 and both Alice and Bob's measurement results. In the second scheme,
   Alice and Bob perform local measurements on their respective particles.
    Bob can infer Alice's secret message  by his measurement results and both Alice and Charlie's encoding
    classical information. In both schemes Alice can deliver secret information to Bob in deterministic manner.
  In the two new quantum secure conditional direct communication schemes,
transmitting one bit secret message, Alice only requires to apply a local unitary transformation or perform a
local measurement on one particle and  to send one bit classical information to Bob for her to transmit one bit
secret message. The operation on one qubit is simpler than that on two qubits. So our two proposed schemes are
even simpler but more economical and more efficient schemes with EPR pairs.

It is worth notice that the classical messages should be authenticated, since a classical channel is not safe:
the message from Charlie to Bob, e.g., could easily be corrupted by Eve, who may want to disturb the
communication between Alice and Bob. Without authentication Bob cannot be sure whether the signals he receives
are indeed from Charlie. In addition, an actual quantum channel will never be perfect. Furthermore, without
quantum repeaters, the quantum channel will connect at most a very restricted spatial distance. Such limitations
and problems will be discussed in the future.

\nonumsection{Acknowledgements} \noindent This work was supported by Hebei Natural Science Foundation of China
under Grant No: A2004000141 and Key Natural Science Foundation of Hebei Normal University.

\nonumsection{References}

\end{document}